\documentclass[journal]{IEEEtran}
\usepackage{graphicx}
\usepackage{epstopdf}

\ifCLASSINFOpdf
\else
\fi
\begin{document}
%
\title{Green Traffic Engineering for Future Core Networks}
%
%
%
%
\author{George Athanasiou\\
Automatic Control Lab, School of Electrical Engineering, \\
KTH Royal Institue of Technology, Sweden, \\
E-mail: \ttfamily{\{georgioa\}@kth.se
}}
\maketitle

\begin{abstract}
An important goal towards the design of Future Networks is to achieve the best ratio of performance to energy consumption and at the same time assure manageability. This paper presents a general problem formulation for Energy-Aware Traffic Engineering and proposes a distributed, heuristic Energy-Aware Traffic Engineering scheme (ETE) that provides load balancing and energy-awareness in accordance with the operator's needs. Simulation results of ETE compared to the optimal network performance confirm the capability of ETE to meeting the needs of Future Networks.
\end{abstract}


%
\IEEEpeerreviewmaketitle

\section{Introduction}
%
%
%
%
\IEEEPARstart{R}{ecently}, \emph{Network Operators} realized their interest in achieving \emph{energy-aware} network operation, and added this objective in their goals list. The successful spreading of broadband access, the consequential increasing number of customers, the versatile spread of services that need to be supported day by day, and last but not least the increasing energy prices raised the demand to provision broadband services more energy effectively. Unfortunately, the today's network infrastructures, namely routers, switches and other network devices, lack effective energy management solutions. \emph{Traffic Engineering (TE)} plays a crucial role in determining the performance and reliability of the network deployments. The prime challenge of TE is to handle unpredictable traffic changes, both, in \emph{capacity demands} and \emph{actual loads}. \emph{Load balancing} and \emph{congestion avoidance} schemes are applied to cope with sudden load changes, vital for reliable service maintenance, and on demand connection management is provided to perform efficient \emph{service provisioning}.

Taking into account the TE objectives, we present a joint problem formulation for \emph{load balancing} with \emph{energy awareness}. We follow a TE problem formulation where the main objective is to \emph{minimize the maximum link utilization} in the network. By maintaining low link utilization, our approach allows the network operators to optimally exploit the capabilities of the existing infrastructure for a longer time and avoid buying new equipment. Therefore, this policy reduces the \emph{Capital Expenditures (CAPEX)}. Furthermore, we \emph{minimize the energy consumption} by turning the \emph{idle} and the \emph{inefficiently utilized links} (lines of the networks cards), into \emph{sleeping mode}. In other words, based on the network conditions and the traffic requests we try to find the optimal set of links that can be turned into sleeping mode. In this way we achieve improved \emph{Operational Expenditures (OPEX)}. The solution of the previous problem formulations leads to improved load balancing and energy-consumption levels in the network. Therefore, they could be used as a performance benchmark. In order to smoothly introduce the aforementioned major issues in real network deployments, we propose a \emph{distributed Energy-Aware Traffic Engineering (ETE)} scheme. \emph{ETE} is "directed" by \emph{low-complexity heuristic algorithms} that are executed in an autonomous manner, through monitoring of the status of the network and making "intelligent" decisions.

TE receives huge attention as one of the most important mechanisms seeking to optimally maintain network performance. The authors in [1] give an overview of the TE approaches that emerged the last years and placed focus on two major issues: quality of service (QoS) and network resilience. A general classification of these traditional-objective TE approaches is: Intradomain and Interdomain [2], MPLS-based and IP-based [3], [4], Offline and Online [5], [6], Unicast and Multicast [7], [8]. The work in this paper is inspired by these traditional TE approaches. Recently, routing, rate adaptation and network control are reconsidered for energy-efficient network operation [9], [10]. Unfortunately, none of these approaches provide a general problem formulation in the direction of "coupling" the traditional TE objectives with the new challenging objectives, like energy-awareness.

The rest of the paper is organized as follows. In section II we describe the problem formulation and the \emph{ETE} scheme. Section III presents the evaluation study and in section IV we summarize our contribution in this paper.

\section{Traffic Enginnering Objectives}
In this section we give a general formulation of the load balancing and the energy efficiency problems for the operator's networks. Then, we present a distributed \emph{Energy-Aware Traffic Engineering scheme} that follows the guidelines provided by the analytical study. We consider a network model, where each ingress router may have traffic demands for a particular egress router or set of routers and assume multiple paths (MPLS tunnels) to deliver traffic from the ingress to the egress routers. Traffic is split among the available paths at the granularity of a flow, to avoid effects that lead to performance degradation [11], and the paths are computed and re-computed offline by the operator.

\subsection{Load balancing oriented problem formulation}
We assume that for each ingress-egress node pair \(i\) the traffic demand is \(T_i\) and multiple paths \(P_i\) could be used to deliver the traffic from the ingress to the egress node. A fraction of the traffic in \(i\), \(x_{ip}\) is routed across path \(p\) (\(p \in P_i\)). In this model, \(L\) is the set of the links in the network, \(IE\) is the set of ingress-egress node pairs, \(e_l\) is the energy consumption of the port connected to link \(l\), \(P_i\) is the set of paths of ingress-egress node pair \(i\), \(T_i\) represents the traffic demands of ingress-egress node pair \(i\), \(a_l\) is a binary variable (equals to 1 if link \(l\) is active, equals to 0 if link \(l\) is "sleeping"), \(u_l\) is the utilization of link \(l\), \(c_l\) is the capacity of link \(l\), \(x_{ip}\) is the fraction of traffic of ingress-egress node pair \(i\) sent through the path \(p\), \(P_l\) is the set of paths that go through link \(l\), \(L_i\) is the set of links that are crossed by the set of paths \(P_i\) and \(E\) is the operator demand related to the desired energy consumption.

We formulate the problem of optimal splitting the traffic caused by each pair \(i \in IE\) along the available paths, assuring that the maximum link utilization (total traffic over active link divided by the link capacity) in the network is minimized (balanced and stable network operation is assured [12]):
\[\begin{array}{c}
\mathop {\min }\limits_{{x_{ip}}} \mathop {\max }\limits_{l \in L} \sum\limits_{i \in IE} {\sum\limits_{p \in {P_i}} {{a_l}\frac{{{x_{ip}}{T_i}}}{{{c_l}}},} } \\
subject\begin{array}{*{20}{c}}
{}
\end{array}to:\begin{array}{*{20}{c}}
{}&{}&{}&{}&{}&{}&{}&{}&{}&{}&{}&{}&{}&{}&{}&{}&{}&{}
\end{array}\\
{x_{ip}} \ge 0,\forall p \in {P_i},\forall i \in IE\\
{c_l} \ge \sum\limits_{i \in IE} {\sum\limits_{p \in {P_l}} {{x_{ip}}{T_i},\forall l \in L} } \\
\sum\limits_{p \in {P_i}} {{x_{ip}} = 1,} \forall i \in IE\\
{a_l} = \{ 0,1\} ,\forall l \in L\\
{x_{ip}} = [0,1],\forall p \in {P_i},\forall i \in IE
\end{array}\]
The constraints ensure that: the fraction of traffic caused by a specific node pair \(i\) sent along a path cannot be negative, the capacity of each link cannot be outreached and the traffic splitting along the available paths meets the traffic demands.

\subsection{Energy consumption oriented problem formulation}
Then, we introduce energy-awareness by identifying the set of links in the network that could be turned into sleeping mode. Therefore, we formulate the problem of finding the optimal set of "sleeping" links in order to achieve minimum energy consumption in the communication (sum of the energy consumption of the active links):
\[\begin{array}{c}
\mathop {\min }\limits_{{a_l}} \sum\limits_{l \in L} {{e_l}{a_l}} ,\\
subject\begin{array}{*{20}{c}}
{}
\end{array}to:\begin{array}{*{20}{c}}
{}&{}&{}&{}&{}&{}&{}&{}&{}&{}&{}&{}&{}&{}&{}&{}&{}&{}
\end{array}\\
{x_{ip}} \ge 0,\forall p \in {P_i},\forall i \in IE\\
{a_l} - {u_l} \ge 0,\forall l \in L\\
{c_l} \ge \sum\limits_{i \in IE} {\sum\limits_{p \in {P_l}} {{x_{ip}}{T_i},\forall l \in L} } \\
{u_l} = \sum\limits_{i \in IE} {\sum\limits_{p \in {P_i}} {\frac{{{x_{ip}}{T_i}}}{{{c_l}}},\forall l \in L} } \\
\sum\limits_{p \in {P_i}} {{x_{ip}} = 1,} \forall i \in IE\\
{a_l} = \{ 0,1\} ,\forall l \in L\\
{x_{ip}} = [0,1],\forall p \in {P_i},\forall i \in IE
\end{array}\]
We have the same constraints here. We also need to ensure that the utilized links cannot be turned into sleeping mode.

The energy consumption of an active link is affected by the maximum rate that can support and it's utilization. In our formulation, the calculation of the energy consumption of link \(l\), \(e_l\), with capacity \(c_l\), is based on a simple model proposed in [13] (used also in several approaches in literature):
\({e_l} = PowerConsumption({c_l}) \times UtilizationFactor(l).\)
\(PowerConsuption(c_l)\) is the base power consumption of link \(l\) with capacity \(c_l\) and the \(UtilizationFactor(l)\) is the scaling factor to account for the utilization of each link.

\subsection{Heuristic Energy-Aware Traffic Engineering mechanism}
In this section we present a distributed heuristic mechanism, which approaches the optimal energy-aware TE solution. The main constituents of the proposed mechanism are the following low-complexity algorithms:
\begin{itemize}
  \item \textbf{Load Balancing (LB)}: \emph{Given the \(a_l\) values for the links in the network, find the corresponding \(x_{ip}\) values that provide balanced network operation in terms of link utilization.}  In order to provide an efficient solution, each ingress node (for each connected egress node) investigates the paths that go through the maximum utilized link. Then, the ingress node "relieves" this link by moving a portion of traffic \(\Delta x\) and provisioning it proportionally to the rest paths (inverse procedure of progressive filling). This procedure continues till convergence to the optimal \(x_{ip}\) values (converge to optimal solution based on [12]).
  \item \textbf{Energy Saving (ES)}: \emph{Given the \(x_{ip}\) values (that resulted) from \textbf{LB}, find the maximum set of links that could be turned into sleeping mode}. Each ingress node finds the routers that are part of the active routes (to the connected egress nodes) and turn the lines of their network card that are not used (by any path) into sleeping mode.
\end{itemize}

\begin{figure}\vspace{-0.2in}
\centering
\includegraphics[width=2.8in]{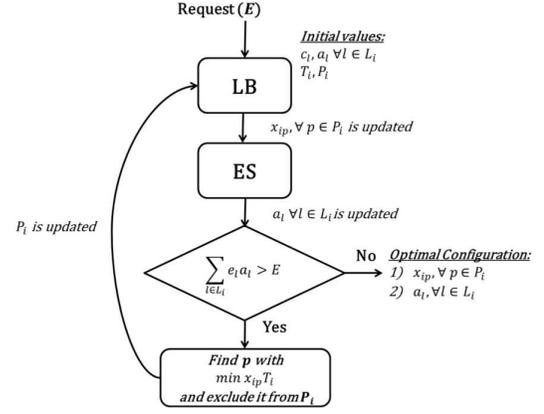}\vspace{-0.1in}
\caption{Heuristic energy-aware load balancing mechanism.} \label{fig1}
\vspace{-0.2in}
\end{figure}

The proposed mechanism (Fig. 1) gets as input the operator's request (\(E\)), as far as the energy consumption is concerned. Then, \textbf{LB} and \textbf{ES} are executed by each ingress node (for each connected egress node) to balance the utilization of the links (that belong to their paths), and turn the non-utilized links into sleeping mode. Next, the new energy consumption level is compared to \(E\) in order to realize if we have reached the desired state. If not, the heuristic mechanism continues by excluding the path \(p\) with the minimum \(x_{ip}T_i\) (lightest path). The heuristic mechanism iterates based on the updated \(P_i\) values, optimizes \(x_{ip}\) and \(a_l\) values \(\forall p \in {P_i}, l \in {L_i}\) and finally, stops when the operator's energy goal is achieved.

Since our mechanism is executed in a distributed manner (by each ingress node), there is no centralized management and coordination of the whole process. The ingress nodes \emph{"independently"} chose a random interval and when this interval expires, \emph{ETE} is executed. In this way we avoid conflicts and we achieve \emph{distributed coordination}. For example, in case that an ingress node puts some traffic to a specific link (increases its utilization) the other ingress nodes that will execute \emph{ETE} are not allowed to turn that link into sleeping mode.

\begin{figure}\vspace{-0.2in}
\centering
\includegraphics[width=3in]{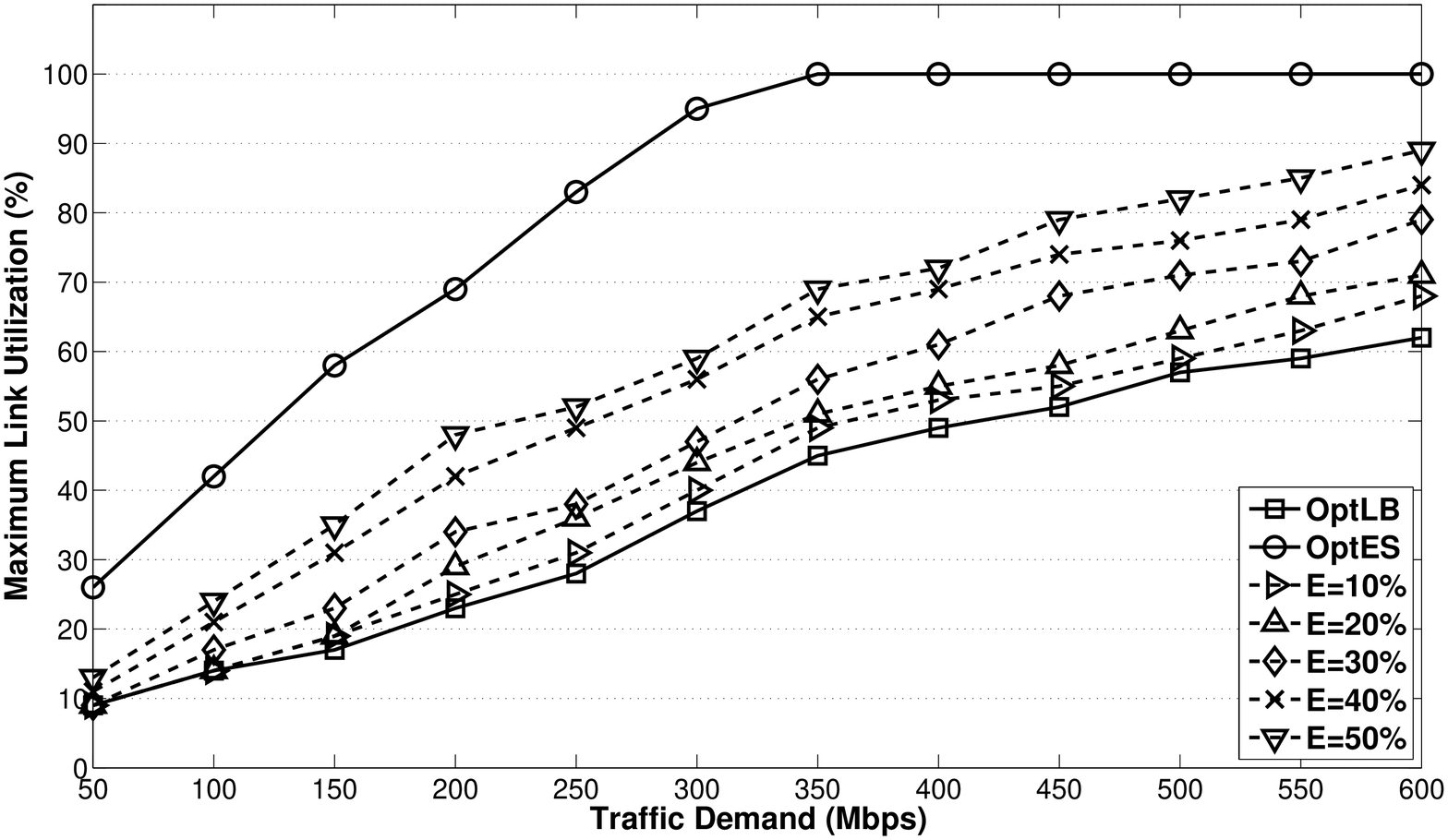}\vspace{-0.1in}
\caption{Maximum link utilization vs. total traffic demand.} \label{fig2}
\vspace{-0.15in}
\end{figure}

\begin{figure}
\centering
\includegraphics[width=3in]{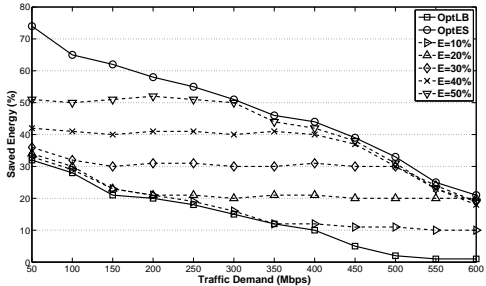}\vspace{-0.1in}
\caption{Percentage of saved energy vs. total traffic demand.} \label{fig3}
\vspace{-0.2in}
\end{figure}

\section{Evaluation}
In this section we present the evaluation study of the proposed scheme. The validation methodology that is adopted uses the optimal solutions as a benchmark in the direction of evaluating \emph{ETE}. We consider a network topology where four ingress nodes send traffic to four egress nodes. Except from these edge nodes, we support 15 core nodes randomly located in a mesh topology. We are using \emph{IBM ILOG CPLEX Optimizer} [14] to find the optimal solutions and evaluate the proposed mechanisms.

Fig. 2 depicts the maximum link utilization in the network vs. the total traffic demands (traffic that must be served in the network). We observe that the performance of \emph{ETE} is bounded by the optimal solutions of the load balancing (OptLB) and the energy saving (OptES) problems and varies based on the operator's demands (e.g. Alg10 represents the performance of \emph{ETE} when \emph{10\%} energy saving is requested i.e. \emph{E=10\%}). The results shown in Fig. 2 prove that in terms of maximum link utilization \emph{ETE} yields solutions close to optimal load balancing.

Fig. 3 depicts the percentage of saved energy vs. the total traffic demands in the network. We observe that the operator's demands are satisfied by \emph{ETE} while ensuring the balanced network operation (close to optimal).

Table I presents simulation results related to the execution of \emph{ETE}. The first column contains the operator's request, the next two show the percentage of the links that are turned into sleeping mode and the routes that are excluded in order to approach the corresponding \emph{E} values. The last column presents the average iterations of \emph{ETE} till convergence.

\begin{table}\vspace{-0.2in}
\caption{ETE Performance} {\footnotesize
\vspace{-0.2in}
\begin{center}
\begin{tabular}{|c|c|c|c|}
  \hline
  \textbf{Requested} & \textbf{Percentage} & \textbf{Percentage} & \textbf{Average ETE } \\
  \textbf{percentage for} & \textbf{of "sleeping"} & \textbf{of routes} & \textbf{iterations till}\\
  \textbf{energy saving} & \textbf{links} & \textbf{excluded} & \textbf{convergence}\\
  \hline
  10\% & 9\% & 3\% & 3  \\
  \hline
  20\% & 20\% & 11\% & 5  \\
  \hline
  30\% & 28\% & 18\% & 7  \\
  \hline
  40\% & 39\% & 24\% & 10  \\
  \hline
  50\% & 48\% & 38\% & 12  \\
\hline
\end{tabular}
\end {center}}
\vspace{-0.2in}
\end{table}

\section{Conclusion}
In this paper we presented a TE analytic approach with main objectives to achieve balanced and energy-efficient network operation. The modeling of these problems inspired the design of an \emph{Energy-Aware Traffic Engineering (ETE)} scheme that tries to meet the requirements of the future networks and pave the way for new "qualified" TE approaches. The simulation results show that \emph{ETE} is capable to achieve performance close to optimal and meet the operator's needs. In other words, \emph{ETE} tends to behave like an optimal load balancer in the network, influenced by the minimum energy saving level that is desired from the operator. \emph{ETE} converges after a small number of iterations, proving in this way it's lightweight operation. Future plans include: extended analytical study, enhancement with learning and autonomic features and implementation.


\begin{thebibliography}{1}

\bibitem{Wang}
N. Wang, K. Ho, G. Pavlou, M. Howarth, An Overview of Routing Optimisation for Internet Traffic Engineering, IEEE Surveys and Tutorials, Vol. 10, No. 1, IEEE, 2008.
\bibitem{Texeira}
R. Teixeira, T. Griffin, A. Shaikh, and G.M. Voelker, Network Sensitivity to Hot-Potato Disruptions, ACM SIGCOMM, August 2004.
\bibitem{Awduche}
Awduche, D.O., MPLS and Traffic Engineering in IP Networks, IEEE Commun. Mag., vol. 37, no. 12, Dec. 1999.
\bibitem{Fortz}
B. Fortz, J. Rexford, M. Thorup, Traffic Engineering with Traditional IP Routing Protocols, IEEE Commun. Mag., vol. 40, no. 10, Oct. 2002.
\bibitem{Goldenberg}
D. K. Goldenberg, L. Qiu, H. Xie, Y. R. Yang, and Y. Zhang, Optimizing Cost and Performance for Multihoming, ACM SIGCOMM 2004.
\bibitem{Elwalid}
A. Elwalid, C. Jin, S. Low, I. Widjaja, MATE: MPLS Adaptive Traffic Engineering, IEEE INFOCOM, 2001.
\bibitem{Kodialam}
M. Kodialam, T.V. Lakshman, Minimum Interference Routing of Applications to MPLS Traffic Engineering, IEEE INFOCOM, 2000.
\bibitem{Kodialam1}
M. Kodialam, T.V. Lakshman, S. Sengupta, Online Multicast Routing with Bandwidth Guarantees: A New Approach Using Multicast Network Flow, IEEE/ACM Trans. Networking, vol. 11, no. 4, Aug. 2003.
\bibitem{Cianfrani}
A. Cianfrani, V. Eramo, M. Listanti, M. Marazza, E. Vittorini, An Energy Saving Routing Algorithm for a Green OSPF Protocol, IEEE INFOCOM 2010, San Diego (USA), 15 March - 19 March 2010.
\bibitem{Nedevschi}
S. Nedevschi, L. Popa, G. Iannaccone, S. Ratnasamy, D. Wetherall, Reducing Network Energy Consumption via Sleeping and Rate-Adaptation , ACM, USENIX, NSDI, 2008.
\bibitem{Kandula}
S. Kandula, D. Katabi, S. Sinha, A. Berger, Flare: Responsive Load Balancing Without Packet Reordering, ACM Computer Communications Review, 2007.
\bibitem{Bertsekas}
D. Bertsekas, R. Gallager, Data Networks. Englewood Cliffs, NJ: Prentice-Hall, 1992.
\bibitem{Mahadevan}
P. Mahadevan, P. Sharma, S. Banerjee, and P. Ranganathan, "A Power Benchmarking Framework for Network Devices," in Proceedings of IFIP Networking, May 2009.
\bibitem{IBM}
IBM ILOG CPLEX Optimizer, http://www-01.ibm.com/software/integration/optimization/cplex-optimizer/
\end{thebibliography}
\end{document}